\documentclass{mem}
\usepackage{natbib}
\usepackage{txfonts}
\usepackage{balance}
\usepackage{graphicx}
\usepackage[a4paper,breaklinks,dvipdfm]{hyperref}
\idline{84}{1}
\usepackage{txfonts} 

\begin{document}

\title{ The effects of flattening and rotation on the temperature of the X-ray
halos of elliptical galaxies
}

   \subtitle{}

\author{
S. \,Posacki, S. \,Pellegrini 
\and L. \,Ciotti
          }

%   \offprints{S. Posacki}

\institute{
Department of Physics and Astronomy, University of Bologna, Viale Berti Pichat
6/2,
I-40127 Bologna, Italy,
\email{silvia.posacki@unibo.it}
}

\authorrunning{Posacki et al.}

\titlerunning{Effects of flattening and rotation on the temperature of X-ray
halos of Es}

\abstract{
Elliptical galaxies have hot coronae with X-ray luminosities and mean gas
temperatures that span over wide ranges. This variation can be partially due to
the energy budget of the hot gas, that depends on the host galaxy structure and
internal kinematics.
With the aid of realistic axisymmetric galaxy models, we performed a
diagnostic study focussed on the effects of galaxy flattening and rotational
support on the hot gas temperature.
}
\maketitle{}

\section{Introduction}
The X-ray observatory $Chandra$ allowed for a study of the hot gas
halos in early-type galaxies (ETGs) with unprecendented quality and
accuracy. With these data, an ETG sample with homogeneous measurements
of the average hot gas temperature and luminosity has recently been
built \citep{boro}, and provides the basis to test the hot
gas properties against other major galaxy properties. For example,
the hot gas content has long been known to be sensitive to the shape
of the mass distribution, and possibly to the mean rotation velocity of
the stars (see \citealt{Pelbook} for a recent review). The same could be
concluded in a recent study focussed on the gas temperature \citep{Pel11}. 
Here we show some preliminary results from an
investigation of the effects of galaxy shape and internal kinematics
on the gas temperature, and on the energy required to extract the gas
from the galaxy potential well, for axisymmetric ETG models.
\section{The galaxy models}
The galaxy models include three mass components: a stellar distribution 
following the approximate ellipsoidal deprojection of the de Vaucouleurs
profile, with axial ratio $0.3~\leq~q~\leq~1$;
a spherically symmetric dark matter (DM) halo, described either by the singular
isothermal sphere (SIS), the Hernquist, the Einasto, or 
the Navarro-Frenk-White profile; and a central massive black hole.
We assume a two-integral phase-space distribution function, so that
$\sigma_{\mathrm R}=\sigma_{\mathrm z}$, and 
the amount of ordered rotational support in the azimuthal direction is
controlled by the \citet{satoh} k-decomposition:
$k=1$ corresponds to the isotropic rotator, whereas for $k=0$ no net rotation is
present, and the galaxy flattening is due to $\sigma_{\varphi}$ only.
The Jeans equations are solved with a numerical code built on purpose, under the
assumption of a constant stellar mass-to-light ratio. All the structural and
dynamical properties are projected face-on (FO) and edge-on (EO), and the
circularized effective radius ($R_{\mathrm e}$) and
the luminosity weighted velocity dispersion within $R_{\mathrm e}/8$
($\sigma_{\mathrm e8}$) are computed. The models are 
forced to follow the most important scaling laws, so that ($R_{\mathrm e}$,
$\sigma_{\mathrm e8}$, $L$) satisfy the Faber-Jackson and
Size-Luminosity relations. % \citep{desroches}.
Finally we calculate fiducial ISM temperatures for the models, as defined in
\citet{Pel11}, under the assumption that the mass sources are
distributed as the stellar density. 
The temperature $T_*= T_{\sigma}+\gamma T_{\mathrm {rot}}$
measures the energy input due to the thermalization of stellar random
($T_{\sigma}$) and ordered ($T_{\mathrm {rot}}$) motions. The actual value of
$\gamma$ is not known a priori, since it depends on the relative
motion between the freshly injected gas and the gas already
permeating the galaxy. It can be obtained only from
hydrodynamical simulations, that suggest it assumes small values (see Negri et
al., this volume). 
The injection temperature $T_{\mathrm{inj}}= T_{*} + T_{\mathrm{SN}}$
takes into account also the contribution of the kinetic energy of SNIa
explosions ($T_{\mathrm{SN}}$). 
Finally, $T_{\mathrm{grav}}$ quantifies the average energy per unit mass
required to extract the gas from the galaxy potential well. 

\section{Results} 
\begin{figure}
\centering
\includegraphics[width=0.475\textwidth]{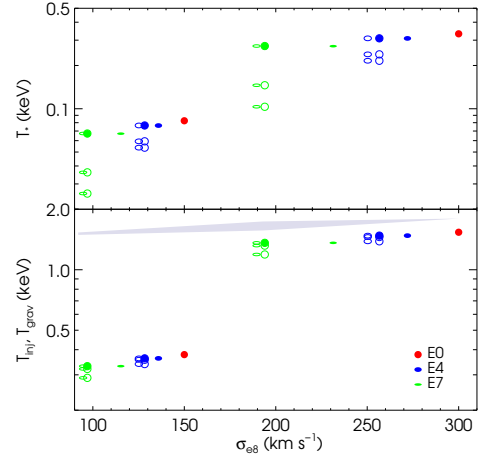} 
\caption{\footnotesize $T_*$ (top) and $T_{\mathrm{grav}}$ (bottom) for two
EO-built sub-families with $\sigma_{\mathrm e8}=150$ and $300\,\mathrm{km\,
s}^{-1}$,
as a function of morphological type and $(k,\gamma)$. Round and elliptical
symbols refer to the FO and EO view of the same models, respectively. Filled
and empty symbols are for $k=0$ and $k=1$. $\gamma=(1,0.5,0)$ decreases
from top to bottom for the empty symbols. The shaded area shows the range
spanned by $T_{\mathrm{inj}}$.}
\label{figT}
\end{figure}

We explored how $T_*$, $T_{\mathrm{inj}}$ and $T_{\mathrm{grav}}$ 
depend on $(q, k, \gamma)$. 
The results presented here refer to a SIS halo truncated at $15\,R_{\mathrm e}$.
The model families are built starting from a
spherical object with luminosity $L$ and effective radius $R_{\mathrm e}$ that
represents the progenitor of a given family. 
The flattened models belonging to a family
are built varying $q$, while keeping both $(L,R_{\mathrm e})$ and the DM halo
fixed. Being $R_{\mathrm e}$
dependent on the line-of-sight (l.o.s.) direction, each family consists of two
sub-families: a FO and an EO-built one.
As a result, $\sigma_{\mathrm e8}$ and the temperatures depend on the choice of
$(q, k, \gamma)$ and l.o.s. direction. 
Figure~\ref{figT} shows $T_*$, $T_{\mathrm{grav}}$ and $T_{\mathrm{inj}}$ for
two EO-built sub-families. 
The models reproduce the observed correlation between 
the mean ISM temperature and $\sigma_{\mathrm e8}$ (e.g. \citealt {boro}), 
an aspect discussed in Posacki et al. (in preparation).
The more the galaxy is flattened, the larger can be the effect 
of rotation: flatter models allow more rotational support, 
so that $T_*$ spans wider ranges as a funtion of $k$ (and consequently of
$\gamma$). 
\normalsize
Instead, $k$ variations have only mild effects on $T_*$ of 
rounder systems. 
Both $T_*$ and $T_{\mathrm{grav}}$ slightly decrease with $q$, and the
$T_{\mathrm{inj}}$ values are dominated by $T_{\mathrm{SN}}$.
If $T_{\mathrm{inj}}>T_{\mathrm{grav}}$ the gas can escape from the 
galaxy (see \citealt{Pel11} for a detailed discussion), thus smaller galaxies
are
more likely to host an outflow/wind region than bigger ones, as already observed
(e.g. \citealt{Pelbook}). 
We found that these temperature trends are almost independent of the specific DM
halo profile.


\begin{thebibliography}{6}
\expandafter\ifx\csname natexlab\endcsname\relax\def\natexlab#1{#1}\fi

\bibitem[{{Boroson} {et~al.}(2011){Boroson}, {Kim}, \& {Fabbiano}}]{boro}
{Boroson}, B., {Kim}, D.-W., \& {Fabbiano}, G. 2011, \apj, 729, 12

% \bibitem[{{Desroches} {et~al.}(2007){Desroches}, {Quataert}, {Ma}, \&
%   {West}}]{desroches}
% {Desroches}, L.-B., {Quataert}, E., {Ma}, C.-P., \& {West}, A.~A. 2007,
% \mnras,
%   377, 402

\bibitem[{{Pellegrini}(2011)}]{Pel11}
{Pellegrini}, S. 2011, \apj, 738, 57

\bibitem[{{Pellegrini}(2012)}]{Pelbook}
{Pellegrini}, S. 2012, in Hot Interstellar Matter in Elliptical Galaxies,
  Springer (Heidelberg), vol. 378, p. 21

\bibitem[{{Satoh}(1980)}]{satoh}
{Satoh}, C. 1980, \pasj, 32, 41
\end{thebibliography}
\end{document}